\documentclass[11pt]{yoshida}

\usepackage{amsmath,amssymb,graphicx,color}

\newcommand{\KUCPlogo}{\hbox{\lower 1.4ex\hbox{
\Huge\boldmath $\cal K$}
\kern -1.15em {\sffamily \bfseries\large\ UCP}}
\kern -4.5em \raise 0.2em\hbox{\lower 1.4ex\hbox{\color{cyan}
\Huge\boldmath $\cal K$}
\kern -1.15em {\color{magenta}\sffamily \bfseries\large\ UCP}
\put(-20,-7){\tiny\it preprint}
}}

\newcommand{\pd}{\partial}

\newcommand{\nn}{\nonumber}
\newcommand{\e}{{\rm e}}

\newcommand{\del}{\delta}
\newcommand{\ra}{\rangle}
\newcommand{\la}{\langle}
\newcommand{\rar}{\rightarrow}

\newcommand{\lra}{\longrightarrow}

\newcommand{\pdm}{\partial_{\mu}}
\newcommand{\pdn}{\partial_{\nu}}

\newcommand{\am}{A_{\mu}}
\newcommand{\an}{A_{\nu}}

\newcommand{\db}{\delta_{\rm B}}
\newcommand{\bdb}{\bar{\delta}_{\rm B}}

\newcommand{\al}{\alpha}

\newcommand{\bfn}{{\bf n}}

\renewcommand{\th}{\theta}

\numberwithin{equation}{section}

\begin{document}

\hbox{\KUCPlogo}

\begin{flushright}

\parbox{3.2cm}{
{KUCP-0175\hfill \\
{\tt hep-th/0012237}}\\
December, 2000
 }
\end{flushright}

\vspace*{2cm}

\begin{center}
 \Large\bf Deconfining Phase Transition in QCD$_4$ and QED$_4$  \\
at  Finite Temperature\\
\end{center}

\vspace*{0.7cm}

\centerline{\large Kentaroh Yoshida}

\begin{center}
{\it Graduate School of Human and Environmental Studies,
\\ Kyoto University, Kyoto 606-8501, Japan. }
\end{center}

\centerline{\tt E-mail:~yoshida@phys.h.kyoto-u.ac.jp}

\vspace*{1.5cm}

\centerline{\bf Abstract}

We investigate the deconfining phase transition in QCD$_4$ and QED$_4$
at finite temperature using a perturbative deformation of topological
quantum field theory (TQFT).
A modified maximal abelian gauge (MAG) is utilized in the analysis.  
In this case, we can derive the linear potential studying the 2D theory through Parisi-Sourlas (PS)
dimensional reduction. 
The mechanism of deconfining phase transition is proposed. It is 
geometrical to discuss the thermal effect on the linear potential. All
we have to do is to investigate the behavior of topological objects
as such instantons and vortices
on a cylinder.  
This is the great advantage of our scenario. This mechanism is also 
applied in the case of QED$_4$. The phase structure at the high temperature of 
QED is investigated using the Coulomb potential on a cylinder. 
It coincides with the result in 
the lattice compact $U(1)$ gauge theory. Also, 
QCD with MAG has the property called the abelian
dominance, which enables us to discuss the deconfinement of
QCD$_4$.

\vspace*{1.5cm}

\thispagestyle{empty}
\setcounter{page}{0}

\newpage

\section{Introduction}

Quark confinement is one of main problems in QCD. This phenomenon is
realized at least in the low energy region (IR region) or at low temperature. 
However, due to asymptotic
freedom the coupling constant becomes large in the IR region, where  
perturbation theory would not be reliable and applicable. 
Therefore, quark confinement should be explained  
 from  non-perturbative aspects of QCD. 
 This phenomenon could be  explained well using the dual super-conductor vacuum
 scenario based on monopole condensation \cite{NM}. However, 
this scenario is not sufficient and many ideas have
 been 
still proposed. One of the recent scenarios, which explains the quark
 confinement in ``{\it continuum}'' QCD, is based on using a perturbative 
deformation of topological quantum field theory
 (TQFT)  \cite{kondo,HT,izawa}. 
This scenario 
describes the quark confinement very well. In particular, if we choose a modified maximal
 abelian gauge (MAG), which is a kind of partial gauge
 fixings \cite{tHooft}, then the linear potential between a static
 quark and antiquark  appears \cite{kondo}. This means the quark confinement in the Wilson criterion  \cite{wilson}.

The properties of QCD medium at finite temperature have been the subject of the intense study. It undergoes a drastic change as the temperature increases. It is believed that confined quarks and gluons are liberated from the certain temperature and the system is deconfined. That is, deconfining phase transition should be caused. 
For high temperature a characteristic energy of quarks and gluons traveling through the medium is  high, and the effective coupling is small. The system presents the quark-gluon plasma (QGP). Since the effective coupling is small, we can use the perturbation theory and the perturbative calculation of its self-energy leads to the thermal mass of gluons and quarks. These mean the Debye screening effect as in the usual plasma.   

 In this paper, we study the deconfining phase  transition of the finite
 temperature 
QCD$_4$ (i.e., thermal QCD) with MAG. In a previous paper \cite{KY}, 
we have investigated the difference of 
the thermal phase transition mechanism between Lorentz type gauge fixing 
and MAG. In  
both cases the TQFT sector can have the phase structure. 
In Lorentz type gauge
fixing this structure is essential to the deconfinement of full QCD$_4$, 
while in the MAG the phase structure of TQFT sector cannot have
relevance in
full QCD$_4$.
Therefore, it has been unclear how we can explain the deconfinement phase 
transition. In this paper, we propose the deconfinement scenario in
thermal QCD$_4$ with MAG, in which all we have to do is to 
investigate the behavior of topological objects such as 
instantons and vortices 
on a cylinder. For example, in an $SU(2)$ QCD$_4$ 
we need to consider the instantons of 2D $O(3)$ non-linear sigma model
(NLSM$_2$ or $\mathbb{C}P^1$ model) on a
cylinder \cite{O3}. However, this theory has asymptotic freedom and their
instanton solution has the size parameter. Therefore the treatment seems
 rather  
difficult because the usual dilute gas sum ansatz is
reliable in the very restricted region.  Hence we concretely  
argue the deconfinement scenario 
in compact QED$_4$ in this paper. In this case, the TQFT sector becomes an $O(2)$ NLSM$_2$
and only to consider their vortices on a cylinder, which have no issues
as the above.  
We also can investigate the behavior at
high temperature using the Coulomb potential on a cylinder. 
Moreover the result can be applied to the abelian
projected effective theory, which is an  abelian gauge theory with asymptotic freedom.
Finally some prospects and issues in studying QCD$_4$  are also discussed.

Our paper is organized as follows. In section 2, we review the method of a
perturbative deformation of the TQFT at zero temperature. We choose a
modified MAG, which have some additional terms and the gauge fixing
term has an $OSp(4|2)$ symmetry.
Quark confinement could be explained analyzing  TQFT sector, 
 in which  the non-perturbative information 
of confinement has been encoded.
This sector is equivalent to a certain 2D theory through 
Parisi-Sourlas dimensional reduction mechanism \cite{PS}, which is caused
due to an $OSp(4|2)$
symmetry.
 We mainly investigate the case of $G=SU(2)$, in which the TQFT
sector becomes an $O(3)$ NLSM on a plane (zero temperature). It 
has also instanton solutions and the linear potential between quark-antiquark 
pair could be induced by the
instanton effect.  In section 3 we consider the finite temperature case. 
The evaluation of 
the rectangular Wilson loop at zero temperature become the evaluation of
two Polyakov loops' correlator. Therefore, 
we can investigate the phase transition from a viewpoint of 2D instantons on a cylinder. 
It is expected that 
the main role in the deconfinement phase 
transition is determined by the behavior of these.
In section 4, we investigate the phase structure of compact QED$_4$. 
The method of a perturbative
deformation can also be applied to QED$_4$ and 
it is known that zero temperature compact QED$_4$ has confining phase at
large coupling  \cite{qed}. In this case, the TQFT sector is an $O(2)$ NLSM$_2$ on 
a cylinder. Of course, at zero temperature it is an $O(2)$ NLSM$_2$ on
2-plane, and a confining potential is induced by the Coulomb gas of
vortices. The phase transition is described by the celebrated 
Berezinskii-Kosterlitz-Thouless (BKT) phase transition \cite{BKT}.
It is natural extension to consider the Coulomb gas on a
cylinder. Thus we can investigate the high temperature region of QED$_4$. 
We conclude in this case that thermal effect shifts  
the value of the coupling in which confining phase transition is caused. 
Moreover, we apply the result of compact QED$_4$ to the scenario 
in  \cite{kondo2} by using the abelian dominance. We could explain the
deconfinement which is induced by the thermal effect.
Section 5 is devoted to conclusions and discussions.

\vspace*{1cm}

\section{QCD$_4$ as Perturbative Deformation of TQFT}

\subsection{Setup}

In this section, we review the method of a perturbative deformation of
TQFT, which is discussed in Refs.\cite{kondo,HT}. Firstly, 
we consider an $SU(N)$ QCD at zero temperature in the 3+1 dimensional  
Minkowskian space-time.  The modification in the case
of the finite temperature system is denoted latter.
We don't include  matter fields, that is,  
consider the gluodynamics. The action for the gauge group $G=SU(N)$ is
\begin{equation}
\label{f2}
 S = -\frac{1}{2g^2} \int d^4 x {\rm Tr}_{G}F_{\mu\nu}F^{\mu\nu}.
\end{equation}
where $F_{\mu\nu}$ is a 
field strength 
\[
F_{\mu\nu} = \pdm\an - \pdn\am -i[\am,\an]
\]
and $SU(N)$ generators  $T^A,~(A=1,\cdots,N^2 - 1)$, which are 
hermite and traceless, are normalized as 
\[
 {\rm Tr}_{G}(T^A T^B) = \frac{1}{2}\del^{AB}. 
\]

In order to construct the quantum field theory from the classical action
(\ref{f2}), it is adequate to use the BRST quantization. Incorporating the 
(anti-) FP ghost field $C(\bar{C})$ and the auxiliary field $B$, we can
construct the BRST transformation $\db$
\begin{eqnarray}
 & & \db \am = D_{\mu}[A]C,~~~~\db C = iC^2, \nn \\
& & ~\db \bar{C} = iB,~~~~~~~~~~~\db B =0, 
\end{eqnarray}
where $D_{\mu}[A]$ is the covariant derivative given by
\[
 D_{\mu}[A] = \pdm -i[\am,~~].
\]
The gauge fixing term can be constructed from the BRST transformation
$\db$ as 
\begin{equation}
 S_{\rm GF + FP} = -i \db\int d^4x G_{\rm GF + FP}[\am, C, \bar{C}, B],
\end{equation}
where $G_{\rm GF + FP}[\am, C, \bar{C}, B]$ is determined by the gauge
fixing 
condition . If we perform the gauge fixing in the Lorentz
gauge in usual manners, then  we have
\begin{equation}
 G_{\rm GF + FP} = {\rm Tr}_{G}\left[\bar{C}(\pdm A^{\mu} + \frac{\al}{2}B)\right] = \bar{C}^{A}\left(\pdm A^{A\mu} + \frac{\al}{2}B^A\right) ,
\end{equation} 
where $\al$ is a gauge parameter.

In this paper, we choose the modified MAG\footnote{This choice is rather
specific, and more general gauge fixing given by 
\[
G_{\rm GF + FP} = {\rm Tr}_{G/H}\left[\am A^{\mu} - i\alpha C\bar{C} \right]
\]
is also possible. In this case, the interesting phenomenon, which is
discussed in Ref.\cite{KS}, is caused.} 
\begin{eqnarray}
 G_{\rm GF + FP}[\am, C, \bar{C}, B] = \bdb {\rm Tr}_{G/H}\left[\am A^{\mu} + 2iC\bar{C}\right] = \bdb \left[\frac{1}{2}\am^a A^{a\mu} + iC^a\bar{C}^a
\right], 
\end{eqnarray}
where $H$ is the maximal abelian subgroup, which for example 
is $U(1)^{N-1}$ in $G=SU(N)$, the subscript $a$ denotes non-diagonal generators and $\bdb$ is the anti-BRST transformation:
\begin{eqnarray}
 & &\bdb \am = D_{\mu}[A]\bar{C},~~~~~~~\bdb C = i\bar{B}, \nn \\
 & &~\bdb \bar{C} = i \bar{C}^2,~~~~\bdb \bar{B}=0,~~~B + \bar{B} = \{C,\bar{C}\}. 
\end{eqnarray}
This gauge is MAG with some additional terms.  In this
case, the gauge fixing term has a special symmetry, $OSp(4|2)$  symmetry,
which is very powerful to investigate the linear potential between
 a quark-antiquark pair.

Note that monopoles also exist in MAG. This existence is ensured by
the fact that 
the homotopy group $\pi_2(G/H) = \pi_1(H)$ is non-trivial. These can 
not exist in the Lorentz type gauge. 

\begin{figure}
 \begin{center}
\includegraphics[width=13cm]{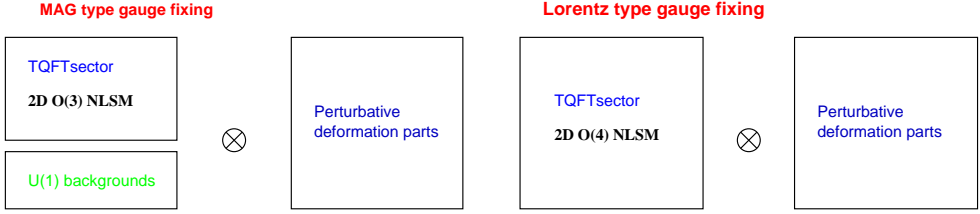}
\caption{\footnotesize The comparison between Lorentz type gauge and 
MAG type gauge 
in $G=SU(2)$.  The phase structure
  of the TQFT sector is preserved in the case of the Feynman type gauge
  fixing. In MAG, however, it is screened by the perturbative fluctuation 
and its information is encoded in the
  $U(1)$ backgrounds. }
\label{comp:fig}
 \end{center}
\end{figure}

Also, modified
Lorentz gauge
\[
 G_{\rm GF + FP}[\am, C, \bar{C}, B] = \bdb {\rm Tr}_{G}\left[\am A^{\mu} + 2iC\bar{C}\right] = \bdb \left[\frac{1}{2}\am^A A^{A\mu} + iC^A\bar{C}^A\right],  
\]
is discussed in Ref.\cite{HT}. The comparison between MAG type and Lorentz type in $G=SU(2)$ 
is shown in Fig.\ref{comp:fig}.

\subsection{Decomposition into Topological Trivial and Non-Trivial Sectors}

In this section, we decompose 
the action of QCD into a topological trivial sector
and a non-trivial sector which is described by a topological quantum
field theory (TQFT), called a TQFT sector. This method was proposed
in  Ref.\cite{HT} in the context of Lorentz type gauge and extended to the
MAG in Ref.\cite{kondo}.
Thus, we may recapitulate QCD
as a perturbative deformation from a TQFT sector. In this method, the
information on the non-perturbative phenomena of QCD, in particular quark confinement and phase structure is encoded in the TQFT
sector. 

We begin with the following decomposition
\begin{equation}
 \am = UV_{\mu}U^{\dagger} + iU\pdm U^{\dagger},
\end{equation}
where we define $\Omega_{\mu}$ as 
\[
 \Omega_{\mu} \equiv iU\pdm U^{\dagger}.
\]
We assume that $\am$ is given by a finite rotation of
$V_{\mu}$. Here $\Omega_{\mu}$ is composed of compact degrees of freedom
$U$ alone,  but $UV_{\mu}U^{\dagger}$ is not compact. In below,
we assume that the non-compact gauge field variable $V_{\mu}$ does
not have topologically non-trivial configuration and all topologically
non-trivial configurations come from the compact gauge group variable
$U$ alone.  As suggested in Ref.\cite{Poly}, it is expected  that
 the perturbation-theoretical study has nothing to do with confinement and 
the real deep
reason for the confinement is encoded into  the topological structure of
the gauge group. 

Secondly, we introduce the FP determinant $\Delta_{\rm FP}[A]$ defined
by 
\begin{equation}
 \Delta_{\rm FP}[A]^{-1} \equiv \int[dU]\prod_{x,A}\del\left(\pd^{\mu}\am^{U^{-1}}(x)\right)
\end{equation}
where $[dU]$ is the gauge invariant Haar measure and so this determinant
is invariant under the gauge transformation, 
\[
 \Delta_{\rm FP}[A] = \Delta_{\rm FP}[A^{U^{-1}}]. 
\]
Then we rewrite the unit as follows, 
\begin{eqnarray}
 1 &=& \Delta_{\rm FP}[A]\int[dU]\prod_{x,A}\del\left(\pd^{\mu}\am^{U^{-1}}\right) \nn \\
&=& \Delta_{\rm FP}[A^{U^{-1}}]\int[dU]\prod_{x,A}\del\left(\pd^{\mu}\am^{U^{-1}}\right) \nn \\
&=& \Delta_{\rm FP}[V]\int[dU]\prod_{x,A}\del\left(\pd^{\mu}V_{\mu}^A\right) \nn \\
&=& \int [dU][d\gamma][d\bar{\gamma}][d\beta]\exp\left[i\int d^4x 2{\rm Tr}_{G}(\beta\pd^{\mu}V_{\mu} + i\bar{\gamma}\pd^{\mu}D_{\mu}[V]\gamma )\right]
\label{FP}
\end{eqnarray}
Here, we define the new BRST transformation $\tilde{\delta}_{\rm B}$ as 
\begin{eqnarray}
 & & \tilde{\delta}_{\rm B} V_{\mu} = D_{\mu}[V]\gamma,~~~\tilde{\delta}_{\rm B} \gamma = i\gamma^2, \nn \\
& & ~ \tilde{\delta}_{\rm B}\bar{\gamma} = i\beta,~~~~~~~~~~\tilde{\delta}_{\rm B}\beta =0. 
\end{eqnarray}
By the use of this $\tilde{\delta}_{\rm B}$, eq.(\ref{FP}) can be
rewritten as 
\begin{equation}
 1 = \int[dU][d\gamma][d\bar{\gamma}][d\beta]\exp\left[i\int d^4x\left(-i \tilde{\del}_{\rm B} \tilde{G}_{\rm GF + FP}[V_{\mu},\gamma,\bar{\gamma},\beta]
\right)\right],
\label{1}
\end{equation}
where $\tilde{G}_{\rm GF + FP}[V_{\mu},\gamma,\bar{\gamma},\beta]$ is
written as 
\begin{equation}
 \tilde{G}_{\rm GF + FP}[V_{\mu},\gamma,\bar{\gamma},\beta] \equiv 2{\rm Tr}_{G}(\bar{\gamma}\pd^{\mu}V_{\mu}).
\end{equation}
When we insert eq.(\ref{1}) into the partition function, it is rewritten as 
\begin{eqnarray}
 Z[J] &=& \int [dU][dC][d\bar{C}][dB]\int[dV][d\gamma][d\gamma]\exp\Bigg[i\int d^4 x \Big(- \frac{1}{2g^2} {\rm Tr}_{G}\left(F_{\mu\nu}[V]F^{\mu\nu}[V]\right)   \nn \\ 
& &-i \tilde{\del}_{\rm B}\tilde{G}_{\rm GF+FP}[V_{\mu},\gamma,\bar{\gamma},\beta] -i \db G_{\rm GF + FP}[\Omega_{\mu} + UV_{\mu}U^{\dagger}, C,\bar{C},B] 
\Big) + i S_{\rm J} \Bigg] 
\end{eqnarray}
where $S_{\rm J}$ is a source term given by
\begin{equation}
 S_{\rm J} = \int d^4x {\rm Tr}_{G}\left[J^{\mu}\left( \Omega_{\mu} + UV_{\mu}U^{\dagger}\right) + J_{\rm c}C + J_{\bar{C}}\bar{C} + J_{\rm B} B \right]. \nn
\end{equation}

Note that the transformation law of $U$ and $V_{\mu}$ under $\db$ and $\bdb$ is
expressed as 
\begin{eqnarray}
 & & \db U = iCU,~~~~\bdb U = i\bar{C}U, \nn \\
 & & \db V_{\mu} = 0,~~~~~~\bdb V_{\mu} = 0. 
\end{eqnarray}
By the use of these transformation laws, we can rewrite $\db G_{\rm GF +
FP}[\Omega_{\mu} + UV_{\mu}U^{\dagger}]$ as 
\begin{eqnarray}
& & \int d^4x \db G_{\rm GF + FP}[\Omega_{\mu} + UV_{\mu}U^{\dagger},C,\bar{C},B]  \nn \\
&=& \int d^4x \db G_{\rm GF + FP}[\Omega_{\mu},C,\bar{C},B] + \int d^4x \left( V_{\mu}^{A}{\mathcal M}^{A\mu}[U] + \frac{1}{2}V_{\mu}^{A}V^{B\mu}{\mathcal K}^{AB}[U]\right),
\end{eqnarray}
where ${\mathcal M}^{A}_{\mu}[U],$ and $ {\mathcal K}^{AB}[U]$  are defined as 
\[
{\mathcal M}^{A}_{\mu}[U] \equiv \db\bdb\left((U T^AU^{\dagger})^a
\Omega^{a}_{\mu}\right),~~~~{\mathcal K}^{AB}[U] \equiv \db\bdb \left((U T^AU^{\dagger})^a(U T^BU^{\dagger})^a  \right). 
\]
Finally, we obtain the following expression of the partition function
\begin{eqnarray}
 Z[J] &=& \int[dU][dC][d\bar{C}][dB]\exp\left(iS_{\rm TQFT} 
+ iW[U;J^{\mu}] \right. \nn \\
& & + \left. i\int d^4x (J^{\mu}\Omega_{\mu} + J_{\rm C}C + J_{\rm \bar{C}}\bar{C} + J_{\rm B} B
\right)
\label{2.16}
\end{eqnarray}
where $S_{\rm TQFT}$ is defined by 
\begin{eqnarray}
 S_{\rm TQFT} &\equiv& -i \db G_{\rm GF + FP}[\Omega_{\mu}, C, \bar{C}, B] \nn \\
&=& -i \int d^4 x\db\bdb {\rm Tr}_{G/H}\left[\Omega_{\mu}\Omega^{\mu} + 2iC\bar{C}\right]
\end{eqnarray}
Here, $W[U;J^{\mu}]$ in eq.(\ref{2.16}) denotes a perturbative
deformation 
and is defined
as 
\begin{eqnarray}
 \exp\left(iW[U;J^{\mu}]\right) \equiv \int[dV][d\gamma][d\bar{\gamma}][d\beta]\exp\bigg(iS_{\rm pQCD}[V_{\mu},\gamma,\bar{\gamma},\beta]  \nn \\
+ i\int d^4x \left(V_{\mu}{\mathcal J}^{\mu} - \frac{1}{2}V_{\mu}^AV^{B\mu}{\mathcal K}^{AB}[U]  \right)\bigg), 
\end{eqnarray}
where ${\mathcal J}_{\mu}$ is the new source term redefined as 
\[
 {\mathcal J}^{A}_{\mu} \equiv U^{\dagger}J^{A}_{\mu}U - {\mathcal M}^{A}_{\mu}[U],
\]
and ${\mathcal K}$ describes the interaction between pQCD and TQFT
sectors. pQCD denotes the
perturbative QCD (topological trivial sector).
Here, the action $S_{\rm pQCD}$  is defined by
\begin{equation}
 S_{\rm pQCD}[V_{\mu},\gamma,\bar{\gamma},\beta] \equiv \int d^4x\left(-\frac{1}{2g^2}F_{\mu\nu}[V]F^{\mu\nu}[V] -i \tilde{\del}_{\rm B}\tilde{G}_{\rm GF + FP}[V_{\mu},\gamma,\bar{\gamma},\beta]\right).
\end{equation}
Note that ${\mathcal M}_{\mu}^{A}[U]$ and ${\mathcal K}^{AB}[U]$ are
interactions between pQCD and TQFT sector.

The perturbative deformation $W[U;J^{\mu}]$ is the generating functional of the
connected Green function of $V_{\mu}$, which describes the perturbative
deformation sector. This should be calculated by the use of the ordinary
perturbation theory with the expansion of  the coupling constant $g$
\begin{eqnarray}
 & & iW[U;J^{\mu}] \equiv {\rm ln}\left\la \exp\left(i\int d^4x[V_{\mu}^A{\mathcal J}^{A\mu} - V_{\mu}^AV^{B\mu}\mathcal{K}^{AB}  ]\right) \right\ra_{\rm pQCD} \nn \\
&=& \frac{g^2}{2} \int d^4x \int d^4y \left\la V_{\mu}(x)V_{\nu}(y)  \right\ra^{c}_{\rm pQCD}\left({\mathcal J}^{\mu}(x){\mathcal J}^{\nu}(y) - \del^4(x-y)\eta^{\mu\nu}\mathcal{K}^{AB}[U]\right) \nn \\
& & + ~{\rm higher~order~of}~g.
\end{eqnarray}
Therefore, $W[U;J^{\mu}]$ is expressed as a power series in the coupling
constant $g$ and goes to zero as $g\rar 0$. It  turns out that the
full QCD is reduced to the TQFT sector in the vanishing limit of coupling
constant. Thus we can interpret the term $W[U;J^{\mu}]$ as the
deformation from the TQFT sector.

\subsection{Relation between Expectation Values}

\begin{figure}
 \begin{center}
\unitlength 0.6pt
\begin{picture}(293,140)
\put(0,0){\includegraphics{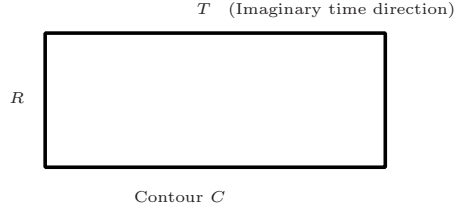}}
\put(17.8,67.7){\makebox(0,0)[lb]{{\tiny $R$}}}
\put(135.9,124.0){\makebox(0,0)[lb]{{\tiny $T$}}}
\put(96.5,5.8){\makebox(0,0)[lb]{{\tiny Contour  $C$}}}
\put(155.6,121.2){\makebox(0,0)[lb]{{\tiny\rm (Imaginary time direction)}}}
\end{picture}
 \end{center}
\caption{\footnotesize The rectangular Wilson loop.}
\label{wil:fig}
\end{figure}

Let us discuss below in the Euclidean metric.
We can define the expectation value in each sector as 
\begin{eqnarray}
 \left\la \mathcal{O}_1 \ldots \mathcal{O}_m\right\ra_{\rm TQFT} &\equiv& \int[dU][dC][d\bar{C}]\mathcal{O}_1 \ldots\mathcal{O}_m \exp\left(- S_{\rm TQFT}[U,C,\bar{C},B] \right),  \\
\left\la \mathcal{O}_1\ldots \mathcal{O}_n\right\ra_{\rm pQCD} &\equiv& \int[dV][d\gamma][d\bar{\gamma}][d\beta]\mathcal{O}_1\ldots\mathcal{O}_n
\exp\left(- S_{\rm pQCD}[V_{\mu},\gamma,\bar{\gamma},\beta]\right),
\end{eqnarray}
and  reconstruct the expectation value of the full QCD$_4$. 
If the inserted operator is decomposed as $f(A) = g(V,U)h(U)$,
then the full expectation value of this is expressed as  
\begin{eqnarray}
\left\la  f(A)\right\ra_{\rm QCD}
= \left\la  \left\la g(V,U) \right\ra_{\rm pQCD} h(U)  \right\ra_{\rm TQFT} = \left\la \left\la g(V,U)h(U) \right\ra_{\rm TQFT}   \right\ra_{\rm pQCD}. 
\label{2.23}
\end{eqnarray}
Thus we can obtain the full expectation value through the above
expectation 
values in each sector.  
Which of expression eq.(\ref{2.23}) should be utilized in calculating the expectation
values
 depends on the case.  
 In fact, the decomposition of the inserted operator is rather difficult
 in non-abelian gauge group. Our  purpose is to investigate the quark
 confinement and so would like to evaluate a non-abelian Wilson loop
\[
 W_C = {\rm Tr}P\exp\left(ie\oint_{C}dx^{\mu}A_{\mu} \right),
\]
where the contour $C$ is the rectangular loop as shown in Fig.\ref{wil:fig}.
In this case, the expectation value of non-abelian Wilson loop 
is mathematically 
decomposed as the above by the use of non-abelian
Stokes theorem \cite{nast,taira}.  
When we consider the case of $G=SU(2)$ for simplicity\footnote{For $G =
SU(N)$, the expression is rather complicated.}, 
it could be expressed as 
\begin{eqnarray}
& &\left\la W_C[A] \right\ra_{\rm QCD} \nn \\
& = & \int d\mu(U) \left\la \left\la \exp \left[ie \oint_C dx^{\mu} {\rm Tr}\left(\sigma_3 UV_{\mu}U^{\dagger}\right)\right]
 \right\ra_{\rm pQCD} \exp\left[ie \oint_C dx^{\mu} \Omega^3_{\mu}\right] \right\ra_{\rm TQFT} 
\label{NAST}
\end{eqnarray}
where $\Omega^3_{\mu} \equiv 2 {\rm Tr}(T^3\Omega_{\mu})$ and $d\mu$ is
the invariant Haar measure of the coset space $SU(2)/U(1)$. 
When we expand eq.(\ref{NAST}) perturbatively, it becomes  
\begin{eqnarray}
& & \left\la W_C[A]\right\ra_{\rm QCD} = \int d\mu(U) \left\la \exp \left[ie \oint_C dx^{\mu}\Omega_{\mu}^3\right] \right\ra_{\rm TQFT} - \frac{e^2}{2}\oint dx^{\mu}\oint dy^{\nu} D_{\mu\nu}(x-y) \times \nn \\
& &\times\int d\mu (U) \left\la \exp\left[ie \oint_C dx^{\mu}\Omega_{\mu}^3\right]2{\rm Tr}\{T^3U(x) T^AU^{\dagger}(x)\}2{\rm Tr}\{T^3U(y)T^AU^{\dagger}(y)\} \right\ra_{\rm TQFT} \nn \\
& & +~{\rm higher~order~of~}e.
\label{nast}
\end{eqnarray}
where we used the following relations
\[
 \left\la V_{\mu}^A(x) \right\ra_{\rm pQCD} = 0,~~~\left\la
 V_{\mu}^{A}(x)V_{\nu}^{B}(y)\right\ra_{\rm pQCD} = \frac{\del^{AB}\del_{\mu\nu}}{4\pi^2 |x-y|^2} \equiv \del^{AB} D_{\mu\nu}(x-y).
\]
Here $e$ is a charge of the external source and proportional to $g$, for example  $e = g/2$ for the fundamental representation. Hence, the above expansion is about the power of $g$. 
As we will see later, the evaluation of the ``{\it abelian}'' Wilson loop
\begin{equation}
 \exp\left[ ie \oint_C dx^{\mu} \Omega_{\mu}^3  \right]
\end{equation}
leads to the linear potential. Note that eq.(\ref{nast}) does not imply the
abelian dominance, which is well known feature in the MAG, but tell us what we should evaluate.

However, in the case of QED$_4$
 the decomposition as the above is completely done as follows \cite{qed}, 
\begin{equation}
 \left\la W_C[A] \right\ra_{\rm QED} = \left\la W_C[\Omega]   \right\ra_{\rm TQFT} \left\la W_C[V]    \right\ra_{\rm pU(1)} 
\end{equation} 
where 
\[
  A_{\mu} = V_{\mu} + \frac{i}{g}U\pdm U^{\dagger},~~~\Omega_{\mu}
  \equiv \frac{i}{g}U\pdm U^{\dagger}.
\]
This is similar with the decomposition of the partition function
\[
 Z = Z_{\rm inst}\cdot Z_{\rm pU(1)}
\]
in the result of Polyakov's work \cite{Poly}, in which 
the linear potential is derived from $Z_{\rm
inst}$ though the theory is on 
3D Euclidean.

\subsection{PS Dimensional Reduction to 2D Theory}

Parisi-Sourlas mechanism can dimensionally reduce 4D
TQFT sector to 2D 
theory \cite{PS}. This is because we have chosen
the special gauge fixing that has an $OSp(4|2)$ symmetry. Though the 
2D space can be taken arbitrarily,  
we should take a 2-plane at zero temperature and a
cylinder at finite temperature  as 2D space respectively 
in order to evaluate
 Wilson loop to derive the linear
potential and investigate quark confinement . 
As a result, the action of TQFT sector
becomes as follows, 
\begin{equation}
\label{coset}
S_{\rm TQFT} = \frac{2\pi}{g^2}\int d^2x {\rm Tr}_{G/H}\left[\Omega_{\mu}\Omega^{\mu} + 2iC\bar{C}\right].
\end{equation}
It describes a coset $G/H$ chiral model on 2D space. 
 
In the  case of $SU(2)$, this action can be rewritten as 
the action of $O(3)$ NLSM,
\begin{equation}
\label{nlsm}
 S_{\rm TQFT} = \frac{\pi}{g^2}\int d^2x \pdm {\bf n}\cdot\pd^{\mu}{\bf n},~~~~{\bf n}\cdot{\bf n} = 1.
\end{equation}
Here we  used the Euler angle representation of 
 an $SU(2)$ matrix 
\begin{eqnarray}
 U(x) &=& \exp\left(i\chi(x)\frac{\sigma_{3}}{2}\right)\exp\left(i\theta(x)\frac{\sigma_{2}}{2}\right)\exp\left(i\varphi(x)\frac{\sigma_{3}}{2}\right) \nonumber \\
&=& \left(
\begin{array}{cc}
\exp\left(\frac{i}{2}(\varphi +\chi )\right)\cos\frac{\theta }{2} & \exp\left( -\frac{i}{2}(\varphi -\chi )\right)\sin\frac{\theta }{2} \\
-\exp\left(\frac{i}{2}(\varphi -\chi )\right)\sin\frac{\theta }{2} & \exp\left( -\frac{i}{2}(\varphi +\chi )\right)\cos\frac{\theta }{2}
\end{array}
\right),
\end{eqnarray} 
and  parameterized an unit length vector field as
\begin{equation}
{\bf  n}(x)\equiv 
\left(
\begin{array}{c}
n^{1}(x) \\
 n^{2}(x) \\
 n^{3}(x)
\end{array}
\right)
\equiv 
\left(
\begin{array}{c}
\sin\theta(x)\cos\varphi(x) \\
\sin\theta(x)\sin\varphi(x) \\
\cos\theta(x) 
\end{array}
\right).
\end{equation}

We can investigate a TQFT sector through 
an NLSM$_2$. In particular, the expectation values of both theories are given as follows,
\begin{equation}
 \left\la \mathcal{O}_1\ldots \mathcal{O}_n \right\ra_{\rm TQFT_4} =  \left\la \mathcal{O}_1\ldots \mathcal{O}_n \right\ra_{\rm NLSM_2}.
\end{equation}

\subsection{Confinement and Static Potential}

Here, we can  evaluate the
abelian Wilson loop through $O(3)$ NLSM instantons as follows,  
\begin{eqnarray}
 & &\left\la W_{C} [\Omega] \right\ra_{\rm TQFT_4} \equiv\left\la \exp \left[ie \oint_C dx^{\mu}\Omega_{\mu}^3(x)\right]\right\ra_{\rm TQFT_4} \nn \\
&=& \left\la \exp \left[2\pi i \left(\frac{e}{g}\right) Q_{{\rm NLSM}_2}\right]\right\ra_{\rm TQFT_4} = \left\la \exp \left[2\pi i \left(\frac{e}{g}\right) Q_{{\rm NLSM}_2}\right]\right\ra_{\rm NLSM_2}
\end{eqnarray}
where $e$ is a charge of an external source, and 
\begin{eqnarray}
Q_{{\rm NLSM}_2} = \frac{1}{8\pi}\int_S d^2x \epsilon_{\mu\nu}\bfn(x)\cdot (\pdm\bfn(x)\times \pdn\bfn(x)), 
\end{eqnarray}
is an instanton density of an NLSM$_2$. Thus, we can calculate this abelian
Wilson loop by the use of dilute gas approximation\footnote{We could improve the instanton calculation including the interaction between (anti-)instantons \cite{inst}.}, and the result is given as 
\begin{equation}
 \left\la W_C[\Omega] \right\ra_{\rm TQFT_4} = \exp\left(-\sigma A\right).
\end{equation}
Here $A=RT$ is the area spanned by the contour $C$,  and $\sigma$, which
 is a string tension of confining string, 
is 
 given by
\begin{equation}
\label{2.36}
 \sigma = 2B\e^{-S_1},~~~S_1 = \frac{8\pi}{g^2},
\end{equation}
for the fundamental representation. In eq.(\ref{2.36}) $S_1$ is the 1-instanton action and $B$ is the constant derived from the integration of the instanton
moduli. When the contribution of the perturbative deformation part is
included, the full Wilson loop expectation value is written as 
\begin{equation}
 \left\la W_C[A]  \right\ra_{\rm QCD} = \e^{-\sigma RT} \left[1 + \left(\frac{3}{4}\right)\frac{e^2}{4\pi R} T f(R) + \cdots\right],
\end{equation}
and the full static potential between a pair of quark and antiquark  is expressed by  
\begin{equation}
\label{potential}
 V(R) = \sigma R - \left(\frac{3}{4}\right)\frac{e^2}{4\pi R}f(R) + \cdots,
\end{equation}
where $f(R)$ is a certain function that behaves as $f(R) \rar 1, ( R \rar 0)$.

Thus  the
linear potential is induced  
by the instanton effect in an $O(3)$ NLSM$_2$ in the leading. 
This means quark confinement in the Wilson criterion.  
Hence we must consider the behavior of the instantons 
in order to study the confinement and deconfinement essentially.
Note that the NLSM$_2$ 
instantons should be interpreted as the points that
monopoles' current lines pierce the 2D space  which has 
been chosen in dimensionally reducing TQFT$_4$ sector (for details, see
 Ref.\cite{kondo}), and so  we can say
 that monopoles are essentially relevant to the confinement.

\section{Confining Phase and Deconfining Phase Transition}

\subsection{Finite Temperature}
 
Now we would like to consider the finite temperature system 
coupled to the thermal bath. 
The imaginary time formalism and real time formalism 
\cite{Ume,NS,LW} are  well known 
procedure to investigate a thermal field dynamics (TFD).  
In both cases, gauge fields 
obey the boundary conditions 
\begin{equation}
 \am(-i\beta, {\bf x}) = \am(0,{\bf x})
\end{equation}
for an imaginary time direction. 

The twisted boundary conditions for the gauge
group element $U(x)$ 
\begin{equation}
\label{bou}
 B_{l}: U(-i\beta,{\bf x}) = U(0,{\bf x})\e^{2\pi li/N},~~~~(l=0,\cdots,N-1)
\end{equation}
can be imposed 
using the element of the center. For $l=0$ it is a periodic one.    

The FP determinant at finite temperature is modified as follows, 
\begin{equation}
 1 \equiv \Delta[A]\frac{1}{N}\sum_{l=0}^{N-1}\int_{B_{l}}[dU]\prod_{x,A}\del\left(\pd^{\mu}\am^{U^{-1}}(x)\right).
\end{equation}
Due to the thermal effect it is decomposed into $N$ independent sectors.
In each sector the gauge transformation obeys each boundary
condition given by eq.(\ref{bou}). It is likely to consider that such decomposition
should have something to do with the domain wall, which is related to the
spontaneous discrete symmetry break down, such as the center symmetry
$\mathbb{Z}_{N}$ \cite{Smilga}. This point remains to be unclear.

At finite temperature, the expectation value is decomposed to the sum as 
\begin{eqnarray}
 \left\la f(A) \right\ra_{\rm QCD} \equiv\frac{1}{N}\sum_{i=0}^{N-1}\left\la \left\la g(V,U)\right\ra_{\rm pQCD} h(U) \right\ra_{\rm TQFT}^{(i)} = 
\frac{1}{N}\sum_{i=0}^{N-1}\left\la \left\la g(V,U)h(U) \right\ra_{\rm TQFT}^{(i)}\right\ra_{\rm pQCD}. 
\end{eqnarray}

\subsection{Boundary Conditions in Reduced 2D Theory}

Let us consider the case of $N=2$ concretely.  
In $N=2$ using PS dimensional reduction both
TQFT$_4^{(i)}~(i=0,1)$ 
sectors are described by the field ${\bf n}(x)$ obeying the periodic condition for imaginary time direction \cite{KY}.

Boundary 
conditions of ${\bf n}(x)$ are derived from the following useful relations 
\begin{equation}
 n^{A}(x)T^{A}=U^{\dagger}(x)T^{3}U(x),~~n^{A}(x)=2{\rm Tr}_{G}[U(x)T^{A}U^{\dagger}(x)T^{3}],~~(A=1,2,3).
\end{equation}
We find that ${\bf n}(x)$ is invariant under $U(1)$ transformation
generated by  $T^{3}$ and it   can be rotated by generators
associated with the coset $SU(2)/U(1)$. 
Also, eq.(\ref{coset}) has the following global $SU(2)_{\rm L}$
symmetry\footnote{In the Lorentz type gauge fixing, TQFT sector
becomes 2D $O(4)$ NLSM$_2$, which 
has a global chiral
symmetry $SU(2)_{\rm
L}\otimes SU(2)_{\rm R}$. While in  MAG $SU(2)_{\rm L}$ symmetry only exists.
Also, this system does not have instanton solutions.}
\begin{equation}
 U \longrightarrow Uh, ~~~~\forall~h\in SU(2)_{\rm L}.
\end{equation} 
Then  ${\bf n}(x)$  transforms as 
\begin{equation}
 n^{A}T^{A} \longrightarrow n^{A}(h^{\dagger}T^{A}h)\equiv n^{\prime A}T^{A}
\end{equation}
and we easily see the action of  $h$ on $n^{A}$
\begin{equation}
\label{n}
 n^{A} \longrightarrow n^{\prime A} = \sum_{B=1}^{3} {\rm ad}(h^{\dagger})^{A}_{~B}n^{B}.
\end{equation}
This means that {\bf n}$(x)$ should be  transformed under $SO(3)$ rotation
but it  is invariant by an action of the center $\mathbb{Z}_{2}$ of $SU(2)$.
By the use of eq.(\ref{n}), boundary conditions for  $U(x)$
can be  
translated into
that on field {\bf n}$(x)$
\begin{eqnarray}
 n^{A}(-i\beta, \sigma) &=& \sum_{B=1}^{3}{\rm ad}(g^{\dagger})^{A}_{~B}n^{B}(0, \sigma),~~~~g\in \mathbb{Z}_2 \nn \\
&=& n^{A}(0, \sigma)
\end{eqnarray}
where $\sigma$ is a spatial coordinate of 2D space. Thus, ${\bf n}(x)$ obeys a 
periodic condition. 

In conclusion, 
boundary conditions of $U(x)$ is irrelevant in MAG from
the viewpoint of reduced 2D theory and its contribution to the full
expectation value is same in each sector, i.e.,
\begin{equation}
\left\la f(A) \right\ra_{\rm QCD} =  \frac{1}{2}\sum_{i=0}^{1}\left\la \left\la g(V,U)h(U) \right\ra_{\rm TQFT}^{(i)}\right\ra_{\rm pQCD}  \lra
\left\la \left\la g(V,U)h(U) \right\ra_{\rm TQFT}
\right\ra_{\rm pQCD}.
\label{3.19}
\end{equation}
This point is different
from the case of the Lorentz type where the field variables on a reduced 2D theory
 obey
twisted boundary conditions in each TQFT sector. 

In the case of $N\geq 3$, there are some possibilities to take the
coset space, so it is nontrivial whether similar result is concluded.  At least
if we take the flag space $F_N \cong
SU(N)/U(1)^{N-1}$ (maximal abelian) or the complex projective space
$\mathbb{C}P^{N-1} \cong SU(N)/(SU(N-1)\times U(1))$ as the coset, 
then the similar result 
seems to be followed \cite{taira}.  Due to the $U(1)$ factor in the
coset the structure such as eq.(\ref{3.19}) would be followed.

\paragraph{Comment on phase structure of TQFT sector}

If we consider only TQFT sector without perturbative deformation part (or
consider pure gauge configuration only), the twist
factor becomes arbitrary element of $SU(2)$ instead of the center, and more general 
boundary conditions of $U(x)$ are allowed as follows,
\begin{equation}
 B_g:~~U(-i\beta,{\bf x}) = U(0,{\bf x})g,~~~~(\forall~g\in G).
\end{equation}
Also, ${\bf n}(x)$
can obey  twisted boundary conditions. In this case, TQFT sector
can have the phase structure though spontaneously symmetry breaking (SSB), 
though it
is forbidden in 2D theory by the novel Coleman-Mermin-Wagner theorem \cite{CMW}. This is
shown in Ref.\cite{HT} by calculating the effective potential.

\subsection{Polyakov Loop and Confinement at Finite Temperature}

Here, using the imaginary time formalism we consider an $SU(N)$ gauge theory at finite temperature, 
in which the order
parameters for deconfining phase transition are the expectation values of 
the Wilson lines wrapping $k$
times around
the compact time  dimensions $\tau = -ix^0$, 
\begin{equation}
 P_k(\vec{x}) = {\rm Tr} P \exp \left[i\int_0^{k\beta} \!\!\! d\tau A_{\tau}(\tau,\vec{x})\right]
\end{equation}
where $P$ denotes the path-ordering and $P_k(\vec{x}),~(k \in \mathbb{Z})$ are also called Polyakov loops.
The theory has a $\mathbb{Z}_N$ symmetry, which allows us to impose the
twisted boundary conditions 
\begin{equation}
 U(\beta, \vec{x}) = \e^{2\pi li/N}U(0,\vec{x})~~~(l=0,\cdots, N-1)
\end{equation}
for gauge group elements. Under this gauge transformation the Polyakov loop
transforms as 
\begin{equation}
 P_k(\vec{x}) \lra \e^{2\pi kli/N}P_k(\vec{x}).
\end{equation}
Therefore $\la P_k(\vec{x}) \ra = 0$ means that $\mathbb{Z}_N$ symmetry is 
unbroken. A Polyakov loop corresponds to an external quark source (in the
fundamental representation of $SU(N)$). 
The free energy of the system (heat bath) is increased
by adding such a source. If we  write this additional free energy as $F_q$,
the expectation value of a Polyakov loop can be  expressed as 
\begin{equation}
\label{free}
 \la P_k(\vec{x}) \ra \sim \e^{-\beta F_q}.
\end{equation} 
From eq.(\ref{free}), 
$\la P_k(\vec{x})\ra =0$ implies that the free energy cost is infinite and 
an isolated quark can not exist in the theory. While if 
$\la P_k(\vec{x}) \ra \neq 0$, it is finite and  
an isolated quark has finite energy, i.e., the
theory no longer confines. In summary, 
\begin{eqnarray}
 \la P_k(\vec{x}) \ra = 0 &\Longrightarrow&~~~ {\rm confining~phase~and}~\mathbb{Z}_N~{\rm symmetry~is~unbroken}, \nn \\
\la P_k(\vec{x}) \ra \neq 0 &\Longrightarrow&~~~ {\rm deconfining~phase~and}~\mathbb{Z}_N~{\rm symmetry~is~broken}. \nn
\end{eqnarray}  
Thus $\mathbb{Z}_N$ symmetry has been considered to be related to
deconfinement transition. Also, $\mathbb{Z}_N$ symmetry is important to
decide the order of the phase transition.

In general, QCD$_4$ is believed to be confined at low
temperature. Therefore, the correlator of two Polyakov loops  
is expected to behave as 
\begin{eqnarray}
\label{exp}
 \la P_{k}(\vec{R})P_{-k}(0)\ra &\sim& \e^{-\beta F_{q\bar{q}}} \lra 0,~~~
(R = |\vec{R}| \lra \infty), \\
F_{q\bar{q}} &=& \sigma R,~~~(\sigma :~{\rm string~tension}) \nn
\end{eqnarray}
at low temperature. That is, it should show the exponentially decay
law. Also, eq.(\ref{exp}) implies $|\la P\ra|^2 = 0$, and so confinement.

\begin{figure}
 \begin{center}
\unitlength 0.8pt
\begin{picture}(560,194)
\put(0,0){\includegraphics{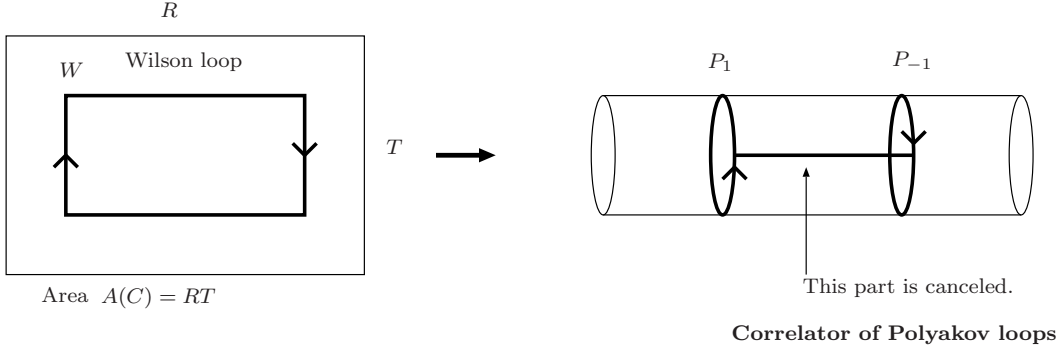}}
\put(104.9,152.3){\makebox(0,0)[lb]{{\scriptsize  }}}
\put(76.8,152.3){\makebox(0,0)[lb]{{\scriptsize Wilson loop}}}
\put(200.5,113.0){\makebox(0,0)[lb]{{\scriptsize $T$}}}
\put(200.5,105.1){\makebox(0,0)[lb]{}}
\put(93.6,177.7){\makebox(0,0)[lb]{{\scriptsize $R$}}}
\put(93.6,169.8){\makebox(0,0)[lb]{}}
\put(397.4,45.5){\makebox(0,0)[lb]{{\scriptsize This part is canceled.}}}
\put(363.6,23.0){\makebox(0,0)[lb]{{\scriptsize\bf Correlator of Polyakov loops}}}
\put(65.5,39.8){\makebox(0,0)[lb]{{\scriptsize $A(C) = RT$}}}
\put(37.4,42.7){\makebox(0,0)[lb]{{\scriptsize Area}}}
\put(37.4,33.1){\makebox(0,0)[lb]{}}
\put(352.4,152.3){\makebox(0,0)[lb]{{\scriptsize $P_1$}}}
\put(439.6,152.3){\makebox(0,0)[lb]{{\scriptsize $P_{-1}$}}}
\put(45.8,149.5){\makebox(0,0)[lb]{{\scriptsize $W$}}}
\end{picture}
 \end{center}
\caption{\footnotesize The correlator of Polyakov loops.  The expectation
  value of a Wilson loop $W$ is equivalent to a correlator of the Polyakov
  loops $P_1$ and $P_{-1}$.}
\label{poly:fig}
\end{figure}

Therefore we would like to evaluate the correlator of the Polyakov loops
in order to study the confinement. Note that the rectangular Wilson loop
at zero temperature becomes the Polyakov loops' correlator at finite
temperature as shown in Fig.\ref{poly:fig}.
Thus, 
\begin{equation}
 \la P_{1}(\vec{R})P_{-1}(0)\ra = \left\la W_C[\Omega] \right\ra 
\end{equation}
is followed. Therefore, in the same way as the case of the zero
 temperature,
 we can derive the linear potential and obtain the following result
\begin{equation}
\label{3.18}
 \la P_1(\vec{R}) P_{-1}(0)\ra \sim \e^{-\beta\sigma R}.
\end{equation}
Note that string tension $\sigma$ depends on the temperature $T$ as 
\begin{equation}
\sigma = 2B(T) \e^{-S_1},~~~B(T)\sim 1/T,~~S_1 = \frac{8\pi}{g^2(T)},
\end{equation}
because the instanton moduli integral is restricted on the cylinder and the effective coupling depends on the temperature. Therefore the string tension depends on the temperature continuously, and decrease as the temperature increases.

Also, a perturbative deformation part would be  modified by the thermal effect.  In particular, we can derive the Yukawa type potential, which means the Debye screening effects as in the usual plasma  
instead of the Coulomb potential at the sufficiently high temperature region. In this time, we use the standard calculation in TFD and use the hard thermal loop (HTL) approximation \cite{LL}. 
Therefore, calculating the contribution of the gluon self-energy to its propagator at 1-loop level with HTL approximation, we obtains the 
\begin{equation}
 \la V_{\mu}^{a}(x)V_{\nu}^b(y)\ra_{\rm pQCD} = \frac{\del_{\mu\nu}\del^{ab}}{4\pi |x-y|}\e^{- m_{D}|x-y|},~~~m_D = \frac{1}{6}g^2 T^2 C_A,
\end{equation}
where $m_D$ is the thermal gluon mass and $C_A = N$ is the group constant.

\subsection{Deconfinement Phase Transition}

The result eq.(\ref{3.18}) is favorable at least in low temperature region but at high temperature region the system should be deconfined. Therefore
it is expected that the string tension should vanish
above a certain temperature, i.e., deconfining phase transition
should  be caused by the thermal effect\footnote{ Unlike in the Lorentz type gauge fixing,  the phase
structure of TQFT sector would not be retained in MAG  once
perturbative deformation part is included. Therefore the deconfining phase
transition must be explained in different way from in the
case of the Lorentz type.}. This problem  has been 
remarked in our previous paper \cite{KY}. 
Let us recall that the
linear potential is
induced by topological objects in our scenario.  
In order for the linear potential to vanish,
the effect of topological objects has to be ignored. For example, a
pair of 
topological objects needs to form the bound state 
and behaves as ``neutral molecules'', or topological objects decay. In such case, the linear potential
would vanish since it is induced by non-zero
topological charge in the rectangular Wilson loop. From the viewpoint of
reduced 2D theory, the thermal effect is realized through the radius of
the 
cylinder. While at low temperature the cylinder is expanded infinitely
and behaves  as 2-plane, at high temperature cylinder collapses and behaves as 
1D line at sufficiently high temperature. 
This implies that it is so simple and geometric to study the 
thermal effect on the linear potential and the deconfining phase transition. 
All we have to do is to investigate the behavior of topological objects on a cylinder. 
This is the great advantage of our deconfinement scenario. 

Thermal effect on the instantons was discussed in Ref.\cite{gross}. At high temperature the cylinder radius constrains the
instanton size and the behavior of these is changed as the 
temperature increases. Large-size instantons are suppressed and so
 instanton effects can be reliably calculated at sufficiently
high temperature. That is, the dilute gas
approximation is rather valid, and so the interaction of instantons should
be suppressed.  In fact, at sufficiently high temperature the instanton effect could be  ignored. 

However, it is difficult to deal with the instanton beyond the dilute gas approximation and to calculate concretely. Therefore we consider
 the case of a compact $U(1)$ gauge theory 
i.e., compact QED without matter. In QCD, we must deal with instanton
 and so worry with the various problem like  infrared problem. But there
 is no problem in 
QED as such.
If the gauge group $U(1)$ is compact, abelian monopoles and 
 the confining phase exist.
In this
case, TQFT sector is an $O(2)$ NLSM on the cylinder and the linear potential is
induced by the vortices of an $O(2)$ NLSM, which should be interpreted as
abelian monopoles from 4 dimensional view.  
Vortices are realized by the
globally neutral Coulomb gas on the cylinder (on 2-plane at zero temperature).
It was showed in Ref.\cite{qed} that the confining-deconfining phase transition at
zero temperature can be described 
by BKT phase transition at certain critical coupling. 
Using the expression of the propagator on the cylinder 
we could find the thermal effect on this
phase transition. In the next section, we will discuss in detail.

\begin{figure}
 \begin{center}
\unitlength 0.8pt
\begin{picture}(548,169)
\put(0,0){\includegraphics{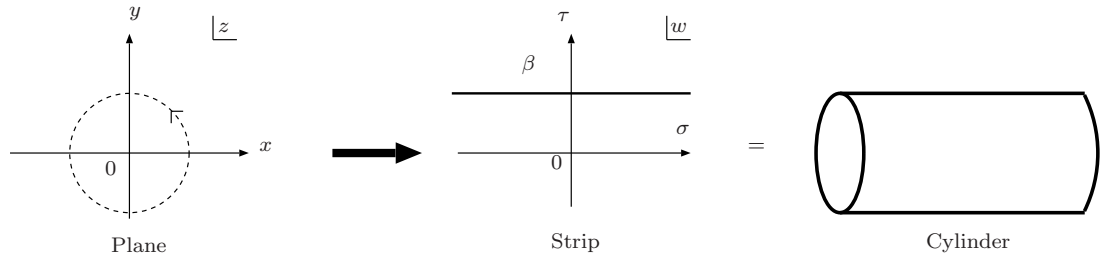}}
\put(362.9,56.6){\makebox(0,0)[lb]{{\scriptsize $=$}}}
\put(270.1,8.8){\makebox(0,0)[lb]{{\scriptsize Strip}}}
\put(447.2,8.8){\makebox(0,0)[lb]{{\scriptsize Cylinder}}}
\put(59.1,45.3){\makebox(0,0)[lb]{{\scriptsize   0}}}
\put(70.4,118.5){\makebox(0,0)[lb]{{\scriptsize $y$}}}
\put(132.2,56.6){\makebox(0,0)[lb]{{\scriptsize $x$}}}
\put(112.6,112.8){\makebox(0,0)[lb]{{\scriptsize $z$}}}
\put(272.9,118.5){\makebox(0,0)[lb]{{\scriptsize $\tau$}}}
\put(329.1,65.0){\makebox(0,0)[lb]{{\scriptsize $\sigma$}}}
\put(256.0,93.2){\makebox(0,0)[lb]{{\scriptsize $\beta$}}}
\put(270.1,48.2){\makebox(0,0)[lb]{{\scriptsize $0$}}}
\put(326.3,112.8){\makebox(0,0)[lb]{{\scriptsize $w$}}}
\put(61.9,8.8){\makebox(0,0)[lb]{{\scriptsize Plane}}}
\end{picture}
\end{center}
\label{cyl:fig}
\caption{\footnotesize Thermal effect is realized
 as circumference of a cylinder or width of a strip. Conformal transformation 
 $w = \frac{\beta}{2\pi}\ln z$ maps the $z$-plane onto the strip.}
\end{figure}

\paragraph{Order of Phase Transition}
Here, we comment on the order of the deconfining phase transition.
There are  two possibilities of the deconfinement mechanism. One is that the string tension vanishes discontinuously because of the vanishing of topological object effect as is discussed above. 
This  corresponds to the first order phase transition.
Thus, the both possibilities of the deconfining phase transition seems to exist.
Another 
is that the string tension continuously vanishes at sufficiently high temperature because of $\e^{-S_1}$ dumping or the $T$-dependence of $B$. This case would correspond to the  second order transition. 

 In general, it is considered that the order of deconfinement phase transition depends on the gauge group. The deconfinement of 4D $SU(N)$ QCD without quarks 
can be  related to the ferromagnetic-paramagnetic transition of $\mathbb{Z}_N$ spin model with the ferromagnetic interactions and $\mathbb{Z}_N$ symmetry plays a crucial role in the
confinement problem. 
In particular, in the case of $N=2$, it is related to 3D $\mathbb{Z}_2$ spin model (Ising), and the deconfinement transition is the second order. Also, in the case of $N=3$ \cite{SY}, it is related to 3D three-state Potts model, the phase transition  of which is the first order one also
expects that a deconfinement phase transition in an $SU(3)$ 
theory, is of first order. 

In the above scenario, it is unclear how we can explain the order of 
deconfinement transition. We could possibly explain it from the intense 
study of the behavior of instantons on reduced 2D theory \cite{DM}. 

\section{Deconfining Phase Transition in Compact QED$_4$}

The phase structure of compact QED$_4$ at zero temperature 
has been worked out in Ref.\cite{qed}.  It has confining phase above the certain
 value of the coupling. Whether such a phase exists or not is decided  
due to the compactness of $U(1)$. If $U(1)$ is
non-compact, such phase can not exist. In next subsection, we study the
thermal effect on the confining phase. 

\subsection{Compact QED$_4$ at Finite Temperature}

We consider the case of compact QED$_4$, which has the confining phase 
in the strong coupling region (UV region).
In the similar way, 
its TQFT$_4$ sector becomes an  $O(2)$ NLSM$_2$, vortices of which 
induce the confining potential. The contribution of vortices to the partition function
is described by 2D classical Coulomb gas.
One of the difference between QCD and QED is that there exists 
no scale parameter (like $\Lambda$ parameter in QCD) in QED  
and also no size parameter in  vortex solutions of $O$(2) NLSM$_2$. 
Hence the treatment is rather simple and strict. 
When we consider the finite temperature case, the TQFT sector becomes
$O$(2) NLSM$_2$ on the cylinder and  vortices contribution could be described by 
the Coulomb gas on the cylinder.

\begin{figure}
 \begin{center}
\unitlength 0.8pt
\begin{picture}(293,101)
\put(0,0){\includegraphics{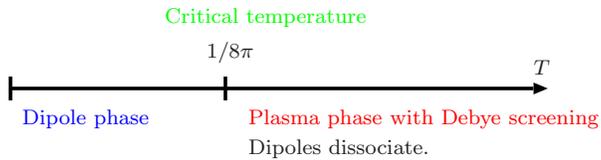}}
\put(104.9,51.2){\makebox(0,0)[lb]{{\scriptsize $1/8\pi$}}}
\put(85.2,68.1){\makebox(0,0)[lb]{{\scriptsize \textcolor{green}{Critical temperature}}}}
\put(259.6,45.6){\makebox(0,0)[lb]{{\scriptsize $T$}}}
\put(6.5,90.6){\makebox(0,0)[lb]{{\scriptsize \textcolor{red}{Phase structure of 2D Coulomb gas}}}}
\put(124.6,20.3){\makebox(0,0)[lb]{{\scriptsize \textcolor{red}{Plasma phase with Debye screening}}}}
\put(124.6,6.2){\makebox(0,0)[lb]{{\scriptsize Dipoles dissociate.}}}
\put(17.8,20.3){\makebox(0,0)[lb]{{\scriptsize \textcolor{blue}{Dipole phase}}}}
\end{picture}
\end{center}
\caption{\footnotesize 2D Coulomb gas has two different phases. Over the
 critical temperature $1/8\pi$, the system is the plasma phase with Debye
 screening, and therefore mass gap exists. Below $1/8\pi$ it is
 the dipoles phase, in which Coulomb charges form dipoles. The system
 has a long range correlation and no mass gap.}
\label{2d-gas:fig}
\end{figure}

The phase structure of 2D Coulomb gas is well known as shown Fig.\ref{2d-gas:fig}. Here, we would like to consider the Coulomb gas on
 the cylinder. 
The propagator on the cylinder is given by
\begin{equation}
\label{cpro}
 G(w - w') = - \frac{1}{4\pi}\ln \left|\e^{\frac{2\pi}{\beta}w} - \e^{\frac{2\pi}{\beta}w'} \right|^2 + \frac{1}{2\beta}{\rm Re}(w + w')
\end{equation}
where $w = \sigma + i \tau$ and $\tau$ is the  imaginary time. We can find eq.(\ref{cpro}) by constructing this propagator on the complex plane and then
mapping it on a cylinder using conformal transformation \footnote{Another derivation of (\ref{cpro}) is given in Appendix A. }. Then,  
requiring that $G(w-w')$ is a function of $w-w'$, an additional term appears.
The propagator eq.(\ref{cpro}) is rewritten as 
\begin{equation}
\label{cpro2}
 G(w - w') = -\frac{1}{4\pi}\ln \left| \e^{\frac{2\pi}{\beta}(w - w')} - 1 \right|^2 + 
\frac{1}{2\beta}{\rm Re}(w - w')
\end{equation}
In the low temperature region ($\beta \rar \infty$), 
\begin{eqnarray}
 G(w - w') &\sim& - \frac{1}{4\pi}\ln \left| \frac{2\pi}{\beta}(w - w')\right|^2 \nn \\
&\sim& - \frac{1}{2\pi}\ln |w - w'|.  
\end{eqnarray}
 As is expected, this eq.(\ref{cpro2})
becomes usual 2D 
Coulomb potential, and  
behaves as the propagator on the plane. 
The constant term can be ignored because of the neutrality of the
Coulomb gas. 
Next, we consider
in the high temperature region ($\beta \rar 0$). In the case of ${\rm Re}(w - w') > 0$,
\begin{eqnarray}
 G(w - w') &\sim& - \frac{1}{4\pi}\ln \left| \e^{\frac{2\pi}{\beta}(w - w')}\right|^2 + \frac{1}{2\beta}{\rm Re}(w - w'), \nn \\
&=& - \frac{1}{2\beta}{\rm Re}(w - w').
\end{eqnarray}
In the case of ${\rm Re}(w - w') < 0$,
\begin{eqnarray}
 G(w - w') &\sim& -\frac{1}{4\pi}\ln | - 1 |^2 + \frac{1}{2\beta}{\rm Re}(w - w'), \nn \\ 
&=& \frac{1}{2\beta}{\rm Re}(w - w').
\end{eqnarray}
Therefore we find in the high temperature region ($\beta\rar 0$) 
\begin{equation}
 G(w - w') \sim -\frac{1}{2\beta}|{\rm Re}(w - w')| = - \frac{1}{2\beta}|\sigma - \sigma'|.
\end{equation}
This is 1D Coulomb potential.
As is expected, the propagator behaves as on a line.
Note that the factor $1/\beta$ appears. This factor does not appear
if we naively start from 1D Coulomb gas. The origin consists in the
finite cylinder radius. This factor is very important in the later discussion.

Let us discuss the behavior of Coulomb gas. It behaves
as 1D Coulomb gas and its partition function is given by 
\begin{equation}
 Z_{1C} = \sum_{n=0}^{\infty}\frac{\zeta^n}{(n!)^2}\int \prod_{j=1}^{n}d\sigma_{j} \exp \left[(2\pi)^2 \frac{1}{\beta g^2}\sum_{i,j}Q_iQ_j |\sigma_i - \sigma_j |\right],~~~\zeta \equiv \e^{- S_{(1)}},
\end{equation}
where $S_{(1)}$ is a single vortex action. Note that  
the temperature of this Coulomb gas system is $\th \equiv \beta g^2 =
g^2/T$.
The 1D Coulomb gas is exactly solvable \cite{1C}. Its
behavior in large $\th$ and small $\th$ is well known. Its phase
structure is very similar with the 2D Coulomb gas. 
For small $\th$
the 1D Coulomb gas behaves as the gas of free ``molecules'',
made up of $+-$ charges pairs bound together. On the contrary, for large
$\th$, the charges are completely deconfined, forming an electrically
neutral ``plasma'' of $2n$ free particles. Now we are considering the high
temperature region ($\beta \rar 0$, that is $T\rar \infty$) i.e., 
the small $\th$, and the 1D Coulomb gas behaves as a gas of 
free ``molecules''. This implies that the linear potential always vanishes
irrelevantly to the definite coupling constant if the temperature is
sufficiently large. Of course, if the coupling
$g^2$ becomes much larger than $T$, that is we consider such an 
energy scale
where $g^2$ becomes too large, the theory is confined again. 
 The relative
measurement between the coupling $g^2$ and the physical temperature $T$ is
important to decide whether confining or deconfining.

\begin{figure}
 \begin{center}
 \includegraphics[height=5cm]{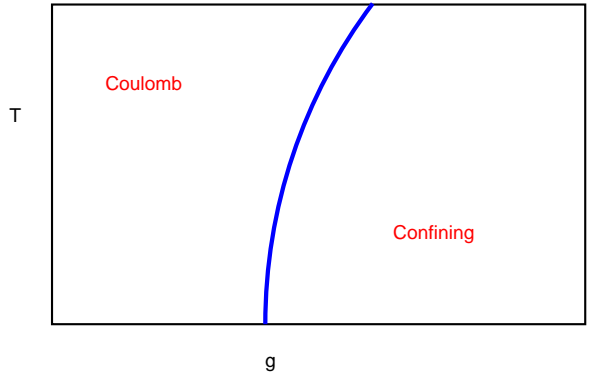}
 \end{center}
\caption{\footnotesize Phase diagram of the lattice compact 
$U(1)$ gauge theory in 4D. The horizontal line is the coupling constant
 and the vertical one is the temperature.}\label{Tg:fig}
\end{figure}

As a result, we could say that compact QED$_4$ 
 is  deconfined at the sufficiently high temperature. 
This nicely corresponds to the lattice compact $U(1)$ gauge theory \cite{SY,SV}.
Its phase diagram is shown in Fig.\ref{Tg:fig}. Therefore we conclude that our scenario describes the confinement-deconfinement in compact QED$_4$ at zero and finite temperature very well.

\subsection{Abelian Dominance and Deconfining Transition}

In this section, using the previous results and the abelian
dominance, we discuss the deconfining phase transition of an $SU(N)$
QCD$_4$.

In MAG, it is expected that off-diagonal gluons are massive and diagonal
gluons are massless and that the diagonal component dominate 
in a sense of Wilsonian effective action. It is confirmed 
by Monte Carlo simulation on a lattice that off-diagonal gluons are
massive  \cite{ad,AS}. As an analytical
derivation has been given at least in the TQFT sector based on the PS 
dimensional reduction to the coset $G/H$ NLSM$_2$  \cite{kondo}.
Once this fact would be assumed, we could deal an $SU(N)$  QCD$_4$ as an 
$U(1)^{N-1}$ abelian gauge theory in the low energy region, which is $N-1$ copies of
compact QED$_4$. That is,  the
low energy effective theory becomes an abelian gauge theory. 
Therefore we
can apply the result of QED$_4$ in previous subsection. 
This scenario has been proposed in Ref.\cite{kondo2}, 
which also can be applied to the finite temperature
 case. This $U(1)^{N-1}$ gauge theory which is derived under the assumption of 
abelian dominance, has asymptotic freedom in contrast to usual (compact)
QED$_4$. Hence the coupling constant
becomes large in the IR region as the same as QCD$_4$. 
 It is clear that the thermal effect
 could cause the deconfinement phase transition at sufficiently high 
temperature from the result of previous subsection if we begin with this low energy effective theory. Hence in this scenario our deconfinement mechanism works well.

In the case of $N=3$ the critical temperature has been estimated in 
Ref.\cite{Hofmann}, based on  the dual Ginzburg-Landau description.

\section{Conclusions and Discussions}

\begin{figure}
 \begin{center}
\includegraphics{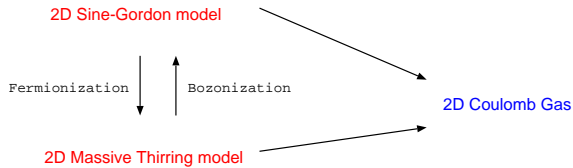}
 \end{center}
\caption{\footnotesize 2D Coulomb gas is intimately 
related to 2D sine-Gordon model and 2D massive Thirring model.}
\label{csm:fig}
\end{figure} 

We have investigated deconfining transition of QCD$_4$ and
QED$_4$ at finite temperature. In our quark confinement scenario, using
PS reduction the analysis is reduced to 2D theory on the cylinder, in
which the thermal effect is realized through the radius of cylinder. 
Therefore, it is very simple and geometrical to investigate thermal
effect 
on the deconfining transition. In order to discuss the deconfinement 
it is enough to study the behavior of topological objects 
on the cylinder with several radii. In particular, we could investigate
 the phase structure of the compact $U(1)$ gauge theory in ``{\it continuum}''. This result agree with the result of lattice.

Taking high temperature limit is equivalent to
compactification on a small circle $S^1$, and so we could discuss the relation
between different dimensions, for example QCD$_4$ and QCD$_5$.  5D theory has been very interesting since it is 
widely discussed in the context of theory with 
extra dimensions. It is interesting work to investigate the connection
between 5D and 4D QCD with the procedure in this paper. Also, our method might be
applied to the system with Higgs fields such as Georgi-Glashow model. In
 Ref.\cite{kogan}, in which 
the Georgi-Glashow model in 3D at finite temperature is discussed,
at high temperature the system become 2D theory and 
BKT phase transition appears. 

It is interesting to consider more general case $G=SU(N)$ rather than
$G=SU(2)$ that is mainly discussed in this paper. If $N \geq 3$, there
is the possibility of choosing the coset. The simplest example is a
maximal torus $SU(N)/U(1)^{N-1}$. In this case, $N-1$ kinds of monopoles 
exist. 
However, a single monopole is enough to induce the linear potential, and so 
$SU(N)/(SU(N-1)\times U(1)) \cong \mathbb{C}P^{N-1}$ is also
possible. The confinement needs at least one $U(1)$ factor in the above
scenario. 
In $\mathbb{C}P^{N-1}$ case, the similar 
argument with the case of $G=SU(2)$ seems to be possible \cite{taira}. 
The large $N$ behavior of $\mathbb{C}P^{N-1}$ model is well
understood. Therefore we might investigate the large $N$ behavior of
QCD$_4$ with our method of the perturbative deformation. 

In this paper, we investigated the phase structure of the compact $U(1)$ gauge 
theory at sufficiently high temperature and strong coupling using  the 
behavior of the 1D Coulomb gas. The intimate relation among the classical 
Coulomb gas, massive Thirring model and sine-Gordon model at zero temperature 
is well known as indicated in Fig.\ref{csm:fig}. This relation at finite temperature is discussed in Refs.\cite{sg,steer}. In particular, we could discuss the phase structure of the compact $U(1)$ gauge theory in terms of sine-Gordon model on the cylinder instead of studying the classical Coulomb gas 
behavior \cite{KY2}. The phase structure of sine-Gordon model at finite 
temperature is discussed in the Ref.\cite{f-sine}.
Also, it is attractive to include dynamical fermions and  consider a 
finite chemical potential, chiral symmetry 
or flavor quantum number $N_{\rm f}$. 

The relation to the result of the dual Ginzburg-Landau theory is also interesting.
In the case of $N=3$ the critical temperature of the deconfinement 
 has been estimated in 
Ref.\cite{Hofmann}, based on  the dual Ginzburg-Landau description.

\vspace*{1cm}
\leftline{\bf\large Acknowledgements}
\vspace*{0.5cm}
\noindent
The author thanks 
S.~Yamaguchi, W. Souma and J.-I.~Sumi 
for useful discussion and valuable comments. He  also would like to
acknowledge 
H.~Aoyama for hospitality in the first stage of his work.

\vspace*{1cm}

\appendix

\leftline{\bf\large Appendix}

\section{The propagator of a periodic boson on a cylinder}
The action 
\begin{equation}
\label{as}
 S = \frac{1}{2}\int_0^{\beta}dx_0\int dx_1~ \pdm\phi \pdm\phi 
\end{equation}
describes massless free boson on a cylinder.

The propagator is given by
\begin{eqnarray}
 D(x-y) &=& \frac{1}{\beta}\sum_{n=-\infty}^{+\infty}\int \frac{dk_1}{2\pi}\e^{-ik\cdot(x-y)}\frac{1}{k^2 + \mu^2}, \nn \\
&=& \frac{1}{\beta}\sum_{n=-\infty}^{+\infty}\int\frac{dk_1}{2\pi}\e^{-ik\cdot(x-y)}\int^{\infty}_{0} ds~ \e^{-s(k^2 + \mu^2)},
\label{a1}
\end{eqnarray}
where $k^2 = k_0^2 + k_1^2, k_0 = 2\pi n/\beta ~(n\in\mathbb{Z})$ is Matsubara frequency. In the above calculation we introduced the mass term $\mu$ in order to avoid the infrared divergence. The above expression (\ref{a1}), using the modified Bessel 
function $K_0(z)$, is rewritten as 
\begin{equation}
 D(x-y) = \frac{1}{2\pi}\sum^{+\infty}_{n=-\infty}K_0\left(\mu\sqrt{(x_0 - y_0 -n\beta)^2 + (x_1 - y_1)^2}\right).
\end{equation}
In the limit $\mu\rightarrow 0$, $D(x-y)$ becomes
\begin{eqnarray}
 D(x-y) &\sim& -\frac{1}{2\pi}\sum_{n=-\infty}^{+\infty}\ln\left(\mu\sqrt{(x_0 - y_0 - n\beta)^2 + (x_1 - y_1)^2}\right), \nn \\
&=& -\frac{1}{2\pi}\ln\left(\mu\beta\sqrt{\cosh \left(\frac{2\pi}{\beta}(x_1 - y_1)  \right) - \cos\left(\frac{2\pi}{\beta}(x_0 - y_0)\right)  }\right).
\end{eqnarray}
Here, we introduce the complex coordinates
\begin{eqnarray}
\label{coor}
 w = x_1 + ix_0,~(\bar{w} = x_1 -i x_0),~~~w'= y_1 +iy_0,~(\bar{w}' = y_1 -iy_0),
\end{eqnarray}
and rewrite $D(x-y)$ by the use of these coordinates (\ref{coor}). Simple 
calculation leads to the following expression, 
\begin{eqnarray}
 D(x-y) = -\frac{1}{4\pi}\ln \left| \e^{\frac{2\pi}{\beta}w} - \e^{\frac{2\pi}{\beta}w'}    \right|^2 + \frac{1}{2\beta}{\rm Re} (w + w') - \frac{1}{4\pi}\ln\left( \frac{1}{2}(\mu\beta)^2\right).
\label{inf}
\end{eqnarray}
The last term in the eq.(\ref{inf}) implies the infrared divergence. 
But this term can be ignored because of the neutral condition of Coulomb gas.
Thus, we obtained the propagator (\ref{cpro}).
 Also, we can relate the sine-Gordon model on the cylinder with the
 classical Coulomb gas on the cylinder, using the propagator
 (\ref{cpro}).

\end{document}